# Practical Attacks on a RFID Authentication Protocol Conforming to EPC C-1 G-2 Standard

MohammadHassan Habibi[1], Mahmud Gardeshi[2], Mahdi R. Alaghband[3]

[1]Faculty of Electrical Engineering, I.H University, Tehran, Iran
`mohamad.h.habibi@gmail.com`
[2]Faculty of Electrical Engineering, I.H University, Tehran, Iran
`mgardeshie2000@yahoo.com`
[3]EEDepartment, Science and Research Campus, Islamic Azad University, Tehran, Iran
`m.alaghband@srbiaui.ac.ir`

*ABSTRACT*

*Yeh et al. recently have proposed a mutual authentication protocol based on EPC Class-1 Gen.-2 standard [1]. They have claimed that their protocol is secure against adversarial attacks and also provides forward secrecy. In this paper we will show that the proposed protocol does not have proper security features. A powerful and practical attack is presented on this protocol whereby the whole security of the protocol is broken. Furthermore, Yeh et al. protocol does not assure the untraceabilitiyand backwarduntraceabilitiy aspects. Namely, all past and next transactions of a compromised tag will be traceable by an adversary.*

## 1. INTRODUCTION

Nowadays Radio Frequency Identification (RFID) technology is incorporated in our daily life. This technology is employed in many applications such as public transportation passes [2], supply chain management [3], e-passport [4], access control systems [5] and etc [6, 7]. RFID systems include tags, readers and back-end server. A tag is a low cost device which has a microchip, small memory and antenna to communicate with the reader. Readers are placed between tags and back-end server as an intermediary for message transmission. On the other hand, the back-end server has the whole information and secret values of all tags.

EPC Class-1 Gen.-2 is a standard that is provided by EPCglobal (Electronic Product Code) organization [8, 9]. This standard provides a framework for RFID communications. EPC C-1 G-2 has restricted tags to some simple arithmetic operations such as CRC (Cyclic Redundancy checksum Code), PRNG (Pseudo Random Number Generator) and bitwise XOR. Therefore, RFID authentication protocols based on EPC C-1 G-2 standard have undergone some difficulties to provide perfect security aspects.

One of the most important challenges related to RFID systems is security problems. In order to have secure authentication protocols, it is important that an adversary does not able to get any information related to the target tag. Privacy and untraceability are two important issues relevant to RFID systems.Thus, an authentication protocol should assure the privacy characteristics such as *untraceability*and *backward untraceability* for tags and their holders [10].

Mentioned attacks aside, different threats related to RFID authentication protocols are mentioned as follows.

• *Information leakage*: When a tag and reader communicate each other, they perform an authentication protocol and exchange some messages. The wireless communication channel between the tag and reader is insecure and it can be eavesdropped by an adversary. Therefore, each authentication protocol should be designed in a way that the adversary, with significant computational capabilities, does not be able to exploit the exchanged messages. Namely, the exchanged messages should not leak any information to the adversary [11].

• *Tag Tracing and tracking*: Tag tracing and tracking are damaging problems in RFID systems. Although the leakage of information is impossible, the untraceability of tag and its holder is not guaranteed in RFID systems. Indeed, untraceability means that if an adversary eavesdrops message transmission between a target tag and a reader at time t, he does not be able to distinguish an interaction of that tag at time t'>t [12].

• *DoS attack*: denial-of-Service (*DoS*) is one category of attacks on RFID systems. An attacker tries to find ways to fail target tag from receiving services. In desynchronization attack, which is one kind of *DoS* attacks, the shared secret values among the tag and the back-end server are made inconsistent by an attacker. Then, the tag and back-end server cannot recognize each other in future and tag becomes disabled [13].

Many RFID authentication protocols have been proposed until now [14, 15, 16, 17, 18, 19, 20, 21, 22, 23, 24]. Although mentioned protocols have wanted to provide secure and untraceable communication for RFID systems; many weaknesses have been found in them[25, 26, 27, 28, 29, 30, 31, 32, 33, 34, 35, 36, 37, 38].However, Yeh et al. [1] have recently proposed a RFID mutual authentication protocol compatible with EPC C-1 G-2 standard that we name SRP (Securing RFID Protocol) in this paper. The authors have claimed that not only does not reveal SRP any information but also it has forward secrecy characteristic. Besides, robustness against *DoS*attack is the other claimed attribute of SRP. In this study, we show that SRP is vulnerable to a powerful and fatal attack that needs only $2^{16}$ off-line PRNG computations. Despite of this attack, the whole security of this protocol will be destroyed inasmuch as RFID system is most vulnerable to tag and reader impersonation and *DoS* attack. Furthermore, we show that the SPR does not assure *untraceability* and *backward untraceability*. The notion *backward untraceability*is defined in section 4.

## 2. RELATED WORK

In this section we briefly study some authentication protocolswhichhave been proposedto provide secure communications in RFID systems.
Dimitriou proposed an RFID authentication scheme that uses a challenge-response mechanism [39].Since the tag identifier remains constant between two successful sessions, this protocol is vulnerable to tracking attacks and tag impersonation attack.
In [40], a lightweight authentication protocol is proposed by Ohkubo et. Al. This scheme provides indistinguishability and forward security characteristics. The scheme is based on a hash chain and uses two dissimilar hash functions *H* and *G*. This protocol does not provide protection against an adversary that tries to de-synchronize the server and the tags, consequently resulting in a DoS attack.

Juels [36] showed that cloning and counterfeiting attacks are applied simply on EPC tags. He proposed an unclonable authentication protocol to solve these problems. However, Duc et al. [20] have presented some weaknesses related to privacy and information leakage in Juels scheme.

In [41],Karthikeyan and Nesterenko suggested a security protocol without complex cryptographic primitives. Only XOR and matrix operations were used in their scheme. Chien and Chen [12] showed that this protocol is vulnerable to replay attacks and does not assure the *untraceability* property.

A mutual authentication protocol under the EPC C-1 G-2 standard was proposed by Chien and Chen [14]. They had used simple XOR, CRC and PRNG in their scheme. In [14] each tag needs to keep an EPC code and two secret keys $K_i$, $P_i$. Secret key $K_i$ is used to tag authentication and secret key $P_i$ is used to reader authentication. Both $K_i$ and $P_i$ are updated in each round whereas *EPC* code is permanent. For each tagsecret values $K_{old}$, $P_{old}$, $K_{new}$, $P_{new}$, *EPC* and *DATA* are stored in database. The protocol is initialed with sending a random number $N_R$ by the reader. As a result, the tag replies with (*M1*, $N_T$) where *M1*=CRC(*EPC*‖$N_R$‖$N_T$)⊕$K_i$. After receiving the tag's response, the database searches for finding the correct tag and its corresponding information ({$K_{old}$, $P_{old}$} or {$K_{new}$, $P_{new}$}). Thenthe database computes *M2*=CRC(*EPC*‖$N_T$)⊕$P_x$ (*x*= *old* or *new*) and sends tag *M2*. At that point the database updates its secret keys as following: $K_{old}$=$K_{new}$, $P_{old}$=$P_{new}$, $K_{new}$=PRNG($K_{new}$) and $P_{new}$=PRNG($P_{new}$). The tag receives *M2* and checks whether *M2*⊕$P_i$=CRC(*EPC*‖$N_T$). If it satisfies, the tag authenticates the database and updates $K_i$ and $P_i$ the same as with the database, else it terminates the protocol.

Lopez et al. [37] showed some weaknesses of Chien and Chen's protocol including tag and reader impersonation and desynchronization attack. They also showed that this protocol does not guarantee forward security and it is vulnerable to tracing attack. Han and Kwon [15] also presented a desynchronization attack and two tag impersonation attacks on Chien and Chen's protocol in new methods. These attacks were mainly based on weak secure properties of CRC.

## 3. REVIEW SRP

### 3.1 Notations

We use the notations the same as the notations were used in the original paper [1].

𝒜:  malicious adversary

$EPC_s$:  16-bit string which is built by XORing six16-bit blocks of EPC code

$C_i$:The database index stored in the tag to find the corresponding record of the tag in the database

$C_{old}$:  The old database index stored in the database

$C_{new}$:  The new database index stored in the database

*DATA*:  The corresponding record for the tag kept in thedatabase

H(.): Hash function

$K_i$: The authentication key stored in the tag for thedatabase to authenticate the tag at the (i + 1)thauthentication phase

$K_{old}$: The old authentication key stored in the database

$K_{new}$: The new authentication key stored in the database

$P_i$: The access key stored in the tag for the tag toauthenticate the database at the (i+1)thauthentication phase

$P_{old}$: The old access key stored in the database

$P_{new}$: The new access key stored in the database

R: the legitimate reader

T :the legitimate tag

X: The value kept as either *new* or *old* to show which keyin the record of the database is found matched withthe one of the tag

$N_Y$: The random number generated by device Y (Y = R or T)

$\gamma_i^{T_i}$: The parameter $\gamma$ related to the tag $T_i$ at time $t_j$

$\oplus$: bitwise XOR

## 3.2 Initialization Phase

For each tag the database is preloaded with nine secretvalues $K_{old}$, $P_{old}$, $C_{old}$, $K_{new}$, $P_{new}$, $C_{new}$, $EPC_S$, $RID$ and $DATA$. Random values $K_0$, $P_0$ and $C_0$ are generated by manufacturer andthe recorded values are set in a way that $K_{old}=K_{new}=K_0$, $P_{old}=P_{new}=P_0$ and $C_{old}=C_{new}=C_0$. Each tag records four values $K_i=K_0$, $P_i=P_0$, $C_i=C_0$ and $EPC_S$ the same as with database.

### 3.3 The (i+1)th Authentication Round

The steps of the authentication phase in the round (*i+1*) of the protocol are presented as follows.

1. The reader generates number $N_R$ at random and sends it to the tag.

2. After receiving $N_R$, first the tag generates random number $N_T$, then it computes:

$$M1 = PRNG\ (EPC_S \oplus N_R) \oplus K_i$$
$$D = N_T \oplus K_i$$
$$E = N_T \oplus PRNG(C_i \oplus K_i)$$

Now the tag forwards ($C_i$, *M1, D, E*) to the reader

3. The reader computes $V=H(RID \oplus N_R)$ and sends $(C_i, M1, D, E, N_R, V)$ to the database.

4. As soon as receiving $(C_i, M1, D, E, N_R, V)$, the database performs the following procedure

(a). For each stored *RID*, the database computes $H(RID \oplus N_R)$ with received $N_R$ and compares the result with *V* to find whether the computed value is the same as with *V*. If the matching is found the database authenticates the reader.

(b). Dependent on the value of $C_i$ one of the following two procedures is occurred:

(i) $C_i = 0$ means it is the first access. For each entry ($K_{old}$, $P_{old}$, $C_{old}$, $K_{new}$, $P_{new}$, $C_{new}$, $EPC_s$, *RID*, *DATA*) the database computes $PRNG(EPC_s \oplus N_R)$, $I_{old}=M1 \oplus K_{old}$ and $I_{new}=M1 \oplus K_{new}$. Then it checks whether $I_{old}$ or $I_{new}$ matches $PRNG(EPC_s \oplus N_R)$. This process is repeated by database until a matching would be found. Dependent on which $I_{old}$ or $I_{new}$ matches, value *X* is set to *old* or *new*.

(ii) If $C_i \neq 0$, the database uses $C_i$ as an index to find the corresponding recorded entry. When the database finds an entry that $C_i$ matches, if it matches $C_{old}$ then the value *X* is set to *old*, otherwise the value *X* is set to *new*. Then corresponding $K_X$ and $EPC_s$ are extracted to check if $PRNG(EPC_s \oplus N_R) \oplus K_X$ is equal to *M1*.
By XORing the extracted $K_X$ with the received D, the database obtains $N_T$ and ensures about correctness of the value $N_T$ by checking whether $N_T \oplus PRNG(C_X \oplus K_X)$ is equal to the received *E*.

(c) Computes $M2=PRNG(EPC_s \oplus N_T) \oplus P_X$ and Info=(DATA$\oplus$RID), and sends them to the reader.

(d) If X = new, it updates the stored values as follows:

$$K_{old} = K_{new} \quad K_{new} = PRNG(K_{new})$$
$$P_{old} = P_{new} \quad P_{new} = PRNG(P_{new})$$
$$C_{old} = C_{new} \quad C_{new} = PRNG(N_T \oplus N_R)$$

But if *X = old*, it just updates $C_{new}$ as $C_{new} = PRNG(N_T \oplus N_R)$.

5. The reader XORs *RID* with the received *Info* and extracts *DATA*, then it sends *M2* to the tag. The tag picks up the stored $P_i$ and computes $P_i \oplus M2$ to find whether it is equal to $PRNG(EPC_s \oplus N_T)$. If the matching would be found, the database is authenticated and the tag updates as follows:

$$K_{i+1} = PRNG(K_i)$$
$$P_{i+1} = PRNG(P_i)$$
$$C_{i+1} = PRNG(N_T \oplus N_R)$$

## 4. VULNERABILITIES of SRP

In this section we will show the most important vulnerabilities of SRP. We first present a practical and powerful attack on SRP in which an adversary obtains the most important secret value of a tag which called $EPC_s$. Aside from the above problem, the SRP is also vulnerable to tracing attacks. We show that the SRP does not provide *backward untraceability* and *untraceability*.

## 4.1 Reveal EP$C_s$

In SRP it is mentioned that $EPC_s$ is a 16-bit string which is constructed from XORing six 16-bit blocks of EPC code. Since $N_R$ and $N_T$ are XORed with $EPC_s$, we conclude the bit lengths of $N_R$ and $N_T$ are the same as bit length of $EPC_s$ Which is 16. Since $K_i$, $P_i$ and $C_i$ are updated by *PRNG*, the bit lengths of them must be equal to the output length of *PRNG* which is 16.

In SRP the bit length of $EPC_s$ is very short and it is also fix in all rounds of the protocol, thus an adversary can exploit this weakness to get $EPC_s$. He just needs to perform two consecutive sessions with the target tag and then calculate $2^{16}$ off-line *PRNG* computations. The procedure of our attack is explained as follows.

- The adversary starts a session with the target tag $T_i$ in the round ($i+1$) by sending random number $N_{R1}$. $T_i$ replies with ($C_i$, $M1_1$, $D_1$, $E_1$). The adversary reserves $M1_1$ and terminates the session. Then he performs the second session with $T_i$ by sending $N_{R2}$ and gets tag's response as ($C_i$, $M1_2$, $D_2$, $E_2$).
- Since the first session is not completed, $T_i$ does not update its secret key $K_i$ for the second session. Hence $M1_1$ and $M1_2$ are constructed as follows:

$$M1_1 = \text{PRNG}\ (EPC_s \oplus N_{R1}) \oplus K_i$$

$$M1_2 = \text{PRNG}\ (EPC_s \oplus N_{R2}) \oplus K_i$$

- A omits $K_i$ by XORing $M1_1$ and $M1_2$:

    $M1_1 \oplus M1_2 = $ PRNG $(EPC_s \oplus N_{R1}) \oplus K_i$ $\oplus$ PRNG $(EPC_s \oplus N_{R2}) \oplus K_i$ = PRNG $(EPC_s \oplus N_{R1}) \oplus$ PRNG $(EPC_s \oplus N_{R2}) = \beta$

    Where $\beta$ is a 16-bit string as a result of $M1_1 \oplus M1_2$.

- Let L=$\{l_1, l_2, \ldots, l_{2^{16}}\}$ be the set of all bit strings with length 16. Since $EPC_s$ is a bit string with length 16, thus $EPC_s \in$ L. By having $\beta$, $N_{R1}$ and $N_{R2}$, the adversary proceeds according to the below algorithm:

    Algorithm 1

    For $1 \leq i \leq 2^{16}$

    Choose $l_i \in$ L

    $\alpha$ =PRNG $(l_i \oplus N_{R1}) \oplus$ PRNG $(l_i \oplus N_{R2})$

    If $\alpha = \beta$ then return $l_i$ as $EPC_s$

    End for

After at most $2^{16}$ execution of the algorithm, the adversary finds the correct $EPC_s$. As a result of the above attack and due to knowing the value of $EPC_s$, we present three important attacks on SRP.

### 4.1.1 Tag Impersonation

By having $EPC_s$, the adversary simply gets the secret key $K_i$ by a passive attack. He listens to the communication channel between the legitimate reader R and the target tag $T_i$ in the round ($i+1$) to obtain $N_{R3}$ and ($C_i$, $M1_3$, $D_3$, $E_3$). Since the adversary has $EPC_s$, he computes PRNG ($EPC_s \oplus N_{R3}$). Thus the secret key $K_i$ is computed as: $K_i = M1_3 \oplus (EPC_s \oplus N_{R3})$ and $K_{i+1} = $ PRNG ($K_i$).

The random number $N_{T3}$ is computed as: $N_{T3} = D \oplus K_i$ and finally the index for the next session is computed as $C_{i+1}$=PRNG ($N_{T3} \oplus N_{R3}$)

Now, the adversary starts a new session with the reader. R sends $N_{R4}$ to him and he replies with ($C_i$, $M1_4$, $D_4$, $E_4$) where $M1_4 = $ PRNG ($EPC_s \oplus N_{R4}) \oplus K_i$, $D_4 = N'_{T4} \oplus K_i$ and $E_4 = N'_{T4} \oplus $ PRNG($C_i \oplus K_i$). Because these values are calculated correctly, the database accepts the adversary and authenticates him.

### 4.1.2 Reader Impersonation and DoS Attack

Aside from tag impersonation, SRP is also vulnerable against two other attacks. By revealing $EPC_s$, the adversary can forge a legitimate reader and then desynchronize the target tag. The procedure of these attacks is explained as following.

- The adversary listens to the communication between R and $T_i$ in the round ($i+1$) to obtain $N_{R5}$, ($C_i$, $M1_5$, $D_5$, $E_5$) and $M2_5$. Since the adversary has $EPC_s$, he computes PRNG($EPC_s \oplus N_{R5}$) and gets the secret key $K_i$ as: $K_i$=$M1_5 \oplus $PRNG($EPC_s \oplus N_{R5}$) and $K_{i+1}$=PRNG($K_i$). The secret key $P_i$ is gotten as: $P_i$=$M2_5 \oplus $PRNG($EPC_s \oplus N_{T5}$) and $P_{i+1}$=PRNG ($P_i$) where $N_{T5}$=$D_5 \oplus K_i$.
- He begins a new session with $T_i$ and sends it $N_{R6}$. $T_i$ replies with ($C_{i+1}$, $M1_6$, $D_6$, $E_6$) which are created with the help of $EPC_s$, $N_{R6}$, $K_i$, $N_{T6}$ and $C_{i+1}$.
- After receiving the tag's response, the adversary extracts $N_{T6}$ as: $N_{T6}$=$D_6 \oplus K_i$, then he computes $M2_5$=PRNG($EPC_s \oplus N_{T6}) \oplus P_{i+1}$ and sends it to the tag.
- $T_i$ checks whether $M2_5 \oplus P_{i+1}$=PRNG($EPC_s \oplus N_{T6}$), since this condition is satisfied, $T_i$ authenticates the adversary and updates its secret values as :

$$K_{i+2} = \text{PRNG}(K_{i+1})$$

$$P_{i+2} = \text{PRNG}(P_{i+1})$$

$$C_{i+2}=\text{PRNG}(N_{R6} \oplus N_{T6})$$

When this session is terminated, the stored secret values on $T_i$ are ($K_{i+2}$,$P_{i+2}$, $C_{i+2}$,$EPC_s$) whereas the database has stored ($K_i$, $P_i$, $C_i$, $K_{i+1}$, $P_{i+1}$, $C_{i+1}$,$RID$,$EPC_s$, $DATA$). Now, they are desynchronized, since the secret values stored in database are completely different from the values stored in the tag.

## 4.2 Privacy Analysis

The authors of SRP have claimed that their protocol has forward secrecy as well as the SRP is resistant to the tracing attacks. We show that the SRP has not forward secrecy. Aside from this weakness, we also present an attacks on *untraceability* of SRP.

### 4.2.1 Privacy Model

Some privacy models have been proposed by researchers to evaluation of RFID protocols [10, 42, 43, 44].Juels and Weis gave a formal definition of the privacy and untraceability model [42]. The samedefinition is described by Ouafi and Phan in their work presented in ISPEC'08 [44] and we will use this model to analyze the SRP protocol.The model that has been described in [44] is summarized as follows.

The protocol parties are tags (T) and readers (R) which interact in protocol sessions. In this model an adversaryAcontrols the communication channel between all parties by interacting either passively or actively with them. The adversaryAis allowed to run the following queries:

- **Execute** (R, T, i )query. This query models the passive attacks. The adversary Aeavesdrops on the communication channel between T and R and gets read access to the exchanged messages between the parties in session i of a truthful protocol execution.
- **Send** (U, V, m, i) query. This query models activeattacks by allowing the adversary$A$to impersonate some reader $U \in R$(respectively tag $V \in T$ ) in some protocol session iand send a messagemof its choice to an instance of some tag $V \in T$(respectively reader $U \in R$ ). Furthermore the adversary A is allowed to block or alert the message mthat is sent from U to V(respectively  V to U) in session iof a truthful protocol execution.
- **Corrupt**(T, $K'$) query. This query allows the adversaryA to learn the stored secretK of the tagT∈T, and which further sets the stored secret to$K'$ .**Corrupt** query means that the adversary has physical access to the tag, i.e., the adversary can read and tamper with the tag's permanent memory.
- **Test** (i, $T_o$, $T_1$) query. This query does not correspond to any of *A*'s abilities, but it is necessary to define the untraceability test. When this query is invoked for session$i$, a random bit $b \in \{0, 1\}$ is generated and then, A is given$T_b \in \{T_o, T_1\}$. Informally, *A*wins if he can guess the bit *b*.

*Untraceable privacy* (*UPriv*) is defined using the game g played between an adversary A and a collection of the reader and the tag instances. The game gisdivided into three following phases:

**Learning phase:**A is given tags $T_o$ and $T_1$ randomly and he is able to send any **Execute**, **Send** and **Corrupt** queries of its choice to $T_0$, $T_1$ and reader.

**Challenge phase:** A chooses two fresh tags $T_0$, $T_1$ to be tested and sends a **Test** (i, $T_o$, $T_1$) query. Depending on a randomly chosen bit $b \in \{0, 1\}$, A is given a tag $T_b$from the set {$T_0$, $T_1$ }.Acontinues making any **Execute**, and **Send** queries at will.

**Guess phase:** finally, A terminates the game g and outputs a bit b' $\in\{0, 1\}$, which is its guess of the value of b.

The success ofAin winning gamegand thus breaking the notion ofUPrivis quantified in termsAadvantage in distinguishing whetherAreceived$T_0$ or $T_1$ and denoted by $\mathbf{Adv}_A^{UPriv}$ (k) where k is the security parameter.

**Adv** $_A^{UPriv}$ (k) =| pr (b = $b'$) − pr (random flip coin) |= | pr (b' = b) - $\frac{1}{2}$ | where

$0 \leq$ **Adv** $_A^{UPriv}$ (k) $\leq \frac{1}{2}$.

In [10], the notion *backward untraceability* is defined as: " *backward untraceability* states that even if given all the internal states of a target tag at time t, the adversary shouldn't be able to identify the target tag's interactions that occur at time t' < t. ".

**4.2.2 Backward Untraceability**

In this section we will show how to break the notion *backward untraceability* in the SRP protocol. Note that $EPC_s$ is fix in the all rounds of SRP and an adversary can exploit this weakness to track a target tag. In particular, consider an adversary A performing the following steps:

**Learning phase:** A sends a **Corrupt**($T_0$, $K'$) query in the round (i+1) and thus obtains $(K_i^{T_0}, P_i^{T_0}, C_i^{T_0}, EPC_{S,i}^{T_0})$.

**Challenge phase:** A chooses two fresh tags $T_0$, $T_1$ to be tested and sends a **Test** (i, $T_o$, $T_1$) query. Depending on a randomly chosen bit b ∈ {0, 1}, A is given a tag $T_b$ from the set {$T_0$, $T_1$}. A makes an **Execute** (R, $T_b$, i) query in the round (i) and as a result, A is given messages $\{N_{R,i-1}^{T_b}, (M1_{i-1}^{T_b}, D_{i-1}^{T_b}, C_{i-1}^{T_b}, E_{i-1}^{T_b})\}$.

**Guess phase:** finally, A terminates the game g and outputs a bit b'∈{0, 1} as its guess of the value of b. In particular, A performs the following procedure to obtain the value b':

He computes $PRNG(EPC_{S,i}^{T_0} \oplus N_{R,i-1}^{T_b}) \oplus M1_{i-1}^{T_b} = \theta$ where $\theta$ is a 16-bit string.

A utilizes the following simple decision rule:

$$b' = \begin{cases} if\ D_{i-1}^{T_b} \oplus E_{i-1}^{T_b} = \theta \oplus PRNG(C_{i-1}^{T_b} \oplus \theta) & b' = 0 \\ otherwise & b' = 1 \end{cases}$$

Thus we have:

**Adv** $_A^{UPriv}$ (k) =| pr ($b' = b$) − pr (random flip coin) | = | pr ($b' = b$) - $\frac{1}{2}$ |= |1 - $\frac{1}{2}$| = $\frac{1}{2}$ ≫ $\varepsilon$

**Proof:** By the fact that $EPC_s$ is a permanent value in the all rounds of the protocol, we have $EPC_{S,i}^{T_0} = EPC_{S,i-1}^{T_0}$. Thus we have the following procedure:

If $T_b = T_0 \Rightarrow PRNG(EPC_{S,i}^{T_0} \oplus N_{R,i-1}^{T_b}) = PRNG(EPC_{S,i}^{T_0} \oplus N_{R,i-1}^{T_0})$ (1)

If $T_b = T_0 \Rightarrow M1_{i-1}^{T_b} = M1_{i-1}^{T_0} = PRNG(EPC_{S,i}^{T_0} \oplus N_{R,i-1}^{T_0}) \oplus K_{i-1}^{T_0}$ (2)

(1), (2) $\Rightarrow$ $PRNG(EPC_{S,i}^{T_0} \oplus N_{R,i-1}^{T_b}) \oplus M1_{i-1}^{T_b} =$ $PRNG(EPC_{S,i}^{T_0} \oplus N_{R,i-1}^{T_0})$ $\oplus$ $M1_{i-1}^{T_0} = PRNG(EPC_{S,i}^{T_0} \oplus N_{R,i-1}^{T_0}) \oplus PRNG(EPC_{S,i}^{T_0} \oplus N_{R,i-1}^{T_0}) \oplus K_{i-1}^{T_0} = K_{i-1}^{T_0} = \theta$ (3)

If $T_b = T_0 \Rightarrow D_{i-1}^{T_b} \oplus E_{i-1}^{T_b} = D_{i-1}^{T_0} \oplus E_{i-1}^{T_0} = N_{T,i-1}^{T_0} \oplus K_{i-1}^{T_0} \oplus N_{T,i-1}^{T_0} \oplus PRNG(C_{i-1}^{T_0} \oplus K_{i-1}^{T_0}) = K_{i-1}^{T_0} \oplus PRNG(C_{i-1}^{T_0} \oplus K_{i-1}^{T_0}) = \theta \oplus PRNG(C_{i-1}^{T_0} \oplus \theta) = \theta \oplus PRNG(C_{i-1}^{T_b} \oplus \theta)$ (4)

### 4.2.3 Untraceability

An authentication protocol for RFID systems should assure the privacy of a tag and its holder. However, many RFID protocols put it at risk by designing protocols where tags answer readers' queries with permanent values, thus performing traceability attacks not only possible but trivial.

Now, we show how the SRP does not guarantee privacy location, thus allowing tags tracking.

**Learning phase:** A sends an **Execute** (R, $T_0$, i+1) query in the round (i+1) by sending $N_{R1}$ and thus obtains $(M1_i^{T_0}, D_i^{T_0}, C_i^{T_0}, E_i^{T_0})$.

**Challenge phase:** A chooses two fresh tags $T_0$, $T_1$ to be tested and sends a **Test** (i+1, $T_o$, $T_1$) query. Depending on a randomly chosen bit b∈ {0, 1}, A is given a tag $T_b$ from the set {$T_0$, $T_1$}. A makes an **Execute** (R, $T_b$, i+1) query by sending $N_{R1}$ and as a result, A is given messages $(M1_i^{T_b}, D_i^{T_b}, C_i^{T_b}, E_i^{T_b})$.

**Guess phase:** finally, A terminates the game g and outputs a bit b' ∈{0, 1} as its guess of the value of b. In particular, A utilizes the following simple decision rule:

$$b' = \begin{cases} \text{if } M1_i^{T_b} = M1_i^{T_0} & b' = 0 \\ \text{otherwise} & b' = 1 \end{cases}$$

Hence we have:

**Adv**$_A^{UPriv}$ (k) = | pr (b' = b) – pr (random flip coin) | = | pr (b' = b) - $\frac{1}{2}$ |= |1 -$\frac{1}{2}$| =$\frac{1}{2}$ ≫ $\varepsilon$

**Proof**: According to the protocol, we have the following equations:

$M1_i^{T_0} = PRNG(EPC_{s,i}^{T_0} \oplus N_{R1}) \oplus K_i^{T_0}$ (5)

$M1_i^{T_b} = PRNG(EPC_{s,i}^{T_b} \oplus N_{R1}) \oplus K_i^{T_b}$ (6)

Note that $T_0$ does not update its secrets in the **Learning phase** and uses the same secret key $K_i$ in both **Learning** and **Challenge phase**. Now we have the following result:

If $T_b = T_0 \Rightarrow M1_i^{T_b} = PRNG(EPC_{s,i}^{T_b} \oplus N_{R1}) \oplus K_i^{T_b} = PRNG(EPC_{s,i}^{T_0} \oplus N_{R1}) \oplus K_i^{T_0} = M1_i^{T_0}$ (7)

# CONCLUSION

In this paper, the significant security flaws in the Yehet al. mutual authentication protocol were showed. We presented a powerful and practical attack on SRP which revealsthe permanent secret value of the target tag. This attack leads to tag and reader impersonation and desynchronization attack on this protocol. Moreover we showed this protocol does not have privacy aspects such as *untraceability* and *backward untraceability*. Our privacy analysis was presented in a formal privacy model.

### ACKNOWLEDGMENT

This work is supported by the Education & Research Institute for ICT, Tehran, Iran.

### REFRENCES

[1] Yeh, T.-C., Wang, Y.-J., Kuo, T.-C., Wang, S.-S.,"Securing RFID systems conforming to EPC Class-1 Generation-2 standard", Expert Systems with Applications 37 (2010) 7678–7683
[2] Transport for London, Oyster card, http://www.oystercard.co.uk.
[3] "Michelin Embeds RFID Tags in Tires", RFID Journal,http://www.rfidjournal.com/article/articleview/ 269 /1/1/. Accessed 17 Jan 2003
[4] Hoepman, J.-H.,Hubbers, E., Jacobs, B., Oostdijk, M., Scherer, R.W.,"Crossing borders: Security and privacy issues of the European e-passport", NAME (IWSEC 2006). LNCS, Springer-Heidelberg, vol. 4266 (2006) 152–167
[5] E.-C. Australia, "Access control, sensor control, and trans-ponders", at: http://www.rfid.com.au/rfid uhf.htm,2008.
[6] D. C. Wyld, "24-Karat protection: RFID and retail jewelry marketing", International Journal of UbiComp (IJU), Vol 1, Num 1, January 2010.
[7]K. K. Khedo, D.Sathan, R.Elaheebocus, R. K. Subramanian, andS.D.V. Rughooputh, "Overlapping zone partitioning localization technique for RFID", International Journal of UbiComp (IJU), Vol 1, Num 2, April 2010.
[8] EPCglobal Inc., http://www.epcglobalinc.org/.
[9] EPCglobal Inc., EPCTM Radio-Frequency Identity Protocols Class-1 Generation-2 UHF RFID Protocols for Communications at 860 MHz – 960 MHz version 1.1.0, Available at [6].
[10] Lim, C.H., and Kwon, T., "Strong and robust RFID authentication enabling perfect ownership transfer", In Proceedings of ICICS '06, LNCS 4307 (2006) 1–20
[11] Van Deursen, T., Radomirovic, S., "Attacks on RFID protocols", Cryptology ePrint Archive, Report 2008/310, 2008. <http://eprint.iacr.org/>.
[12] R. Phan, "Cryptanalysis of a new ultralightweight RFID authentication protocol-SASI", IEEE Transactionson Dependable and Secure Computing 6(4): Oct.-Dec. (2009) 316–320
[13] Peris-Lopez, P., Hernandez-Castro, J.C., Estevez-Tapiador, J.M., and Ribagorda, A., "Vulnerability analysis of RFID protocols for tag ownership transfer", Computer Networks 54 (2010) 1502–1508
[14] Chien, H., Chen, C.,"Mutual Authentication Protocol for RFID Conforming to EPC Class-1 Generation-2 Standards", Computer Standards & Interfaces, 29 (2007) 254–259
[15] Han, D., Kwon, D.: Vulnerability of an RFID authentication protocol conforming to EPC Class-1Generation-2 Standards. Computer Standards & Interfaces 31 (2009) 648–652.
[16] Fu, J., Wu, C., Chen, X., Fan, R., and Ping, L., "Scalable pseudo random RFID private mutual authentication", 2nd IEEE International Conference on Computer Engineering and Technology(ICCET). V. 7, pp. 497-500, China, 2010.

[17] P. Peris-Lopez, J. C. Hernandez-Castro, J. M. Estevez-Tapiador, and A. Ribagorda, "EMAP: An efficient mutual authentication protocol for low-cost RFID tags", In Proc. of IS'06, volume 4277 of LNCS, pages 352–361, Springer-Verlag, 2006.

[18] Gu, Y., Wu, W., "Mutual authentication protocol based on tag ID number updating for low-cost RFID", In Proceedings of the first IEEE International Conference on Network Infrastructure and Digital Content(IC-NIDC2009), pp. 548-551, 2009.

[19] Kim, K. H., Choi, E. Y., Lee, S. M., and Lee, D.H., "Secure EPCglobal Class-1 Gen-2 RFID system against security and privacy problems", In Proc. of OTM-IS'06, volume 4277 of LNCS, pages 362–371. Springer-Verlag, 2006.

[20] Duc, D.N., Park, J., Lee, H., and Kwangjo, K., "Enhancing security of epcglobal Gen-2 RFID tag against traceability and cloning", In Proc. of Symposium on Cryptography and Information Security, 2006.

[21] Li T., and Wang, G., "SLMAP-A secure ultra-lightweight rfid mutual authentication protocol", Proc. of Chinacrypt'07, 2007.

[22] Kulseng, L., Yu, Z., Wei, Y., and Guan, Y., "Lightweight mutual authentication and ownership transfer for RFID Systems", In Proceedings of IEEE INFOCOM 2010, 1-5, CA, March (2010).

[23] Song, B., and Mitchell, C. J., "RFID authentication protocol for low-cost tags", In Wisec 2008, pages 140-147.

[24] Chien, H. Y., "SASI: A new ultralightweightrfid authentication protocol providing strong authentication and strong integrity", IEEE Transactions on Dependable and Secure Computing, 4(4):337–340, 2007.

[25] Peris-Lopez, P., Li, T., Lim, T.-L., Hernandez-Castro, J. C., Estevez- Tapiador, J. M., and Ribagorda, A.,"Vulnerability analysis of a mutual authentication scheme under the epc class-1 generation-2 standard", In Hand. of RFIDSec'08, 2008

[26] Rizomiliotis, P., Rekleitis, E., Gritzalis, S.,"Security analysis of the Song– Mitchell authentication protocol for low-cost RFID tags", Communications Letters, IEEE 13 (4) (2009), pp. 274–276..

[27] Lin, C.-L., and Chang, G.-G., "Cryptanalysis of EPC class 1 generation 2 RFID authentications",Information Security Conference 2007, ChiaYi, Taiwan.

[28] Peris-Lopez, P., Hernandez-Castro, J. C., Estevez-Tapiador, J. M., and Ribagorda, A., "Practical attacks on a mutual authentication scheme under the EPC Class-1 Generation-2 standard", Computer Communications 32 (2009) 1185–1193.

[29] Habibi, M. H.,Gardeshi, M.,Alaghband, M., "Cryptanalysis of a mutual authentication protocol for low-cost RFID", In proceedings of IEEE International Conference on Intelligent Information Networks (ICIIN 2011), UAE, 2011.

[30] Van Deursen, T., Radomiroviˊc, S., "Security of RFID protocols – A case study", Electronic Notes in Theoretical Computer Science 244 (2009) 41–52.

[31] Habibi, M. H., Gardeshi, M., Alaghband, M., "Cryptanalysis of two mutual authentication protocols for low-cost RFID", International Journal of Distributed and Parallel systems, Volume 1, Number 3 ,(2011)

[32] Habibi, M. H., Gardeshi, M., Alaghband, M., "Security analysis of an RFID mutual authentication protocol for RFID systems", In proceedings of IEEE International Conference on Intelligent Information Networks (ICIIN 2011), UAE, 2011.

[33] Habibi, M. H., Gardeshi, M., Alaghband, M., "Attacks and improvements to a new RFID mutual Authentication protocol", In proceedings of Third Workshop on RFID Security: RFIDsec Asia 2011, China, 2011.

[34] Li, T., Deng, R.H., "Vulnerability analysis of EMAP-An efficient RFID mutualauthentication protocol", In: AReS 2007: Second International Conference on Availability, Reliability and Security (2007).

[35] Alomair, B., Lazos, L., and Poovendran, R., "Passive attacks on a class of authentication
protocols for RFID", K.-H. Nam and G. Rhee (Eds.): ICISC 2007, LNCS 4817, pp. 102–115, 2007.

International Journal of UbiComp (IJU), Vol.2, No.1, January 2011

[36] Juels, A., "Minimalist cryptography for low-cost RFID tags", In Proc. of SCN'04, volume 3352 of LNCS, pp. 149–164, Springer-Verlag, 2004.

[37] Peris-Lopez, P., Li, T., Lim, T.-L., Hernandez-Castro, J. C., Estevez- Tapiador, J. M., and Ribagorda, A.,, "Cryptanalysis of a novel authentication protocol conforming to EPC-C-1G-2 standard", Computer Standards & Interfaces, Elsevier Science Publishers, doi:10.1016/j.csi.2008.05.012, 2008.

[38] Han, D., Kwon, D.: Vulnerability of an RFID authentication protocol conforming to EPC Class-1Generation-2 Standards. Computer Standards & Interfaces 31 (2009) 648–652.

[39] Dimitriou, T., "A lightweight RFID protocol to protect against traceability and cloning attacks", In SecureComm, pages 59-66, 2005.

[40] Ohkubo, M., Suzuki, K., Kinoshita, S., "Cryptographic approach to "privacy-friendly tags", In 2003 MIT RFID Privacy Workshop, 2003.

[41] Karthikeyan, S., Nesterenko, M., "RFID security without extensive cryptography", In Proc. of SASN '05, ACM (2005) 63–67.

[42] Juels, A., and Weis, S.A., "Defining strong privacy for RFID", In Proceedings of PerCom '07 (2007) 342–347, http://eprint.iacr.org/2006/137.

[43]Avoine, G., "Adversarial model for radio frequency identification", Cryptology ePrint Archive, report 2005/049. http://eprint.iacr.org/2005/049.

[44] Ouafi, K., and Phan, R.C.-W. "Privacy of recent RFID authentication protocols",L. Chen, Y. Mu, and W. Susilo (Eds.): ISPEC 2008, LNCS 4991, pp. 263–277, 2008.

**Authors**

Mohammad Hassan Habibi received the Bachelor's degree in Telecommunication Engineering From Kerman University, Kerman, Iran, in 2007 and Master's degree in Telecommunication in the field of Cryptography with the honor degree from IHU, Tehran, Iran (2011), where he obtained the Best Student Academic Award. Currently, he is a research assistant (RA) at the research center of cryptography, IHU, Tehran, Iran. His research interest includes: Lightweight cryptography, RFID security, authentication protocols, cryptanalysis, public key cryptography and lightweight primitives.

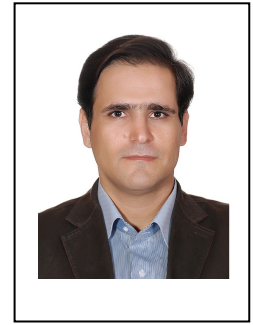

Mahmoud Gardeshi received his Erudition Degree in applied mathematics from Amir Kabiruniversity, Islamic Republic of Iran in 2000. Currently, he is a researcher at the I.H.University. His research interest includes: cryptography and information security

Mahdi R. Alaghband received his B.S. degree in Electrical engineering in 2005 and M.S. degree in Communications, Cryptology & Information Security in 2008. Currently, he is both a Ph.D. candidate at the Department of Electrical and Computer Engineering, Azad University and research assistant on Information Systems and Security Lab (ISSL), EE Dept., Sharif University of Technology. His research interests include authentication protocol especially in RFID systems, security in wireless sensor networks and lightweight cryptographic primitives.

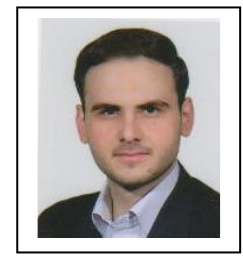